\documentclass[conference]{IEEEtran}
\IEEEoverridecommandlockouts

\usepackage{cite}
\usepackage{amsmath,amssymb,amsfonts, mathtools, amsthm}
\usepackage{algorithmic}
\usepackage{graphicx, color}
\usepackage{textcomp}
\usepackage{xcolor}

\usepackage{threeparttable}
\usepackage{array}
\usepackage{subcaption} 
\usepackage{booktabs}
\usepackage{glossaries}
\usepackage{algorithm}
\usepackage{bbm}
\usepackage{bm}
\usepackage{multirow}

\usepackage[normalem]{ulem}
\usepackage{verbatim}
\usepackage{fancyhdr}
\usepackage{tabularx}
\usepackage{flushend}
\usepackage{balance}
\usepackage{url}

\newacronym{leo}{LEO}{Low Earth Orbit}
\newacronym{isl}{ISL}{Inter-Satellite Link}
\newacronym{fso}{FSO}{free-space optical}
\newacronym{gsl}{GSL}{ground-to-satellite link}
\newacronym{ospf}{OSPF}{Open Shortest Path First}
\newacronym{pan}{PAN}{Path-Aware Networking}
\newacronym{qos}{QoS}{Quality of Service}
\newacronym{e2e}{E2E}{end-to-end}
\newacronym{bgp}{BGP}{Border Gateway Protocol}
\newacronym{ibgp}{iBGP}{interior Border Gateway Protocol}
\newacronym{ebgp}{eBGP}{exterior Border Gateway Protocol}
\newacronym{ntn}{NTN}{Non-Terrestrial Networks}
\newacronym{lsatc}{LSatC}{\gls{leo} Satellite Constellation}
\newacronym{ai}{AI}{Artifical Intelligence}
\newacronym{ip}{IP}{Internet Protocol}
\newacronym{ue}{UE}{User Equipment}
\newacronym{fifo}{FIFO}{First-In First-Out}
\newacronym{snr}{SNR}{Signal-to-Noise Ratio}
\newacronym{ceos}{CEOS}{Conventional Earth Observation Satellites}
\newacronym{ec}{EC}{Edge Computing}
\newacronym{aoa}{AoA}{angle-of-arrivals}
\newacronym{psf}{PSF}{point spread function}
\newacronym{eo}{EO}{Earth Observation}
\newacronym{los}{LOS}{line-of-sight}
\newacronym{gs}{GS}{ground station}
\newacronym{gsd}{GSD}{ground sampling distance}
\newacronym{rf}{RF}{radio frequency}
\newacronym{cdf}{CDF}{cumulative distribution function}

\def\BibTeX{{\rm B\kern-.05em{\sc i\kern-.025em b}\kern-.08em
    T\kern-.1667em\lower.7ex\hbox{E}\kern-.125emX}}
\begin{document}

\title{Impact of Atmospheric Turbulence and Pointing Error on Earth Observation\\
\thanks{All the authors are with the Telecommunications Research Institute (TELMA), Universidad de Málaga, 29071, Málaga, Spain. The work is partially supported by ESA SatNEx V (prime contract no. 4000130962/20/NL/NL/FE), and by the Spanish Ministerio de Ciencia e Innovación under grant PID2022-136269OB-I00 funded by MCIN/AEI/10.13039/501100011033 and “ERDF A way of making Europe”. The view expressed herein can in no way be taken to reflect the official opinion of the European Space Agency. The work of A.M. Mercado-Martínez is also partially supported by Grant DGP\_PRED\_2024\_01603, funded by the Consejería de Universidad, Investigación e Innovación of Junta de Andalucía and the European Union. The work of C. Sánchez-de-Miguel is also partially supported by Grant DGP\_PRED\_2024\_02392, funded by the Consejería de Universidad, Investigación e Innovación of Junta de Andalucía and the European Union. The authors thankfully acknowledge the computer resources, technical expertise and assistance provided by the SCBI (Supercomputing and Bioinformatics) center of the University of Málaga.
}
}
\author{
\IEEEauthorblockN{Celia Sánchez-de-Miguel}
\IEEEauthorblockA{
\textit{TELMA, University of Malaga}\\
Malaga, Spain \\
celiasm@ic.uma.es}
\and
\IEEEauthorblockN{Antonio M. Mercado-Martínez}
\IEEEauthorblockA{
\textit{TELMA, University of Malaga}\\
Malaga, Spain \\
antoniommm@ic.uma.es}
\and
\IEEEauthorblockN{Beatriz Soret~\IEEEmembership{Senior Member,~IEEE}}
\IEEEauthorblockA{
\textit{TELMA, University of Malaga}\\
Malaga, Spain \\
bsoret@ic.uma.es}
\and
\IEEEauthorblockN{Antonio Jurado-Navas~\IEEEmembership{Member,~IEEE}}
\IEEEauthorblockA{
\textit{TELMA, University of Malaga}\\
Malaga, Spain \\
navas@ic.uma.es}
\and
\IEEEauthorblockN{Miguel Castillo-Vázquez~\IEEEmembership{Member,~IEEE}}
\IEEEauthorblockA{
\textit{TELMA, University of Malaga}\\
Malaga, Spain \\
miguelc@ic.uma.es}
}

\maketitle


\begin{abstract}
\gls{eo} imagery is often degraded by atmospheric turbulence and pointing jitter; yet, these effects are rarely considered in datasets used to train AI-based detection models. Based on prior work, this paper presents an enhanced image simulator that enables the incorporation of vertical-path atmospheric turbulence and satellite pointing jitter, arising from platform and sensor vibrations, to generate physically realistic distorted images.
As a case study, vessel detection is evaluated using YOLOv8 and RetinaNet on images generated by the proposed simulator under different levels of turbulence and pointing errors. Results show that YOLOv8 recall decreases from 91\% under ideal conditions to 60\% in the presence of weak turbulence, and falls below 40\% under strong turbulence or jitter. In contrast, RetinaNet demonstrates greater robustness, maintaining approximately 75\% recall across degraded conditions.
These results highlight the importance of incorporating realistic physical degradations into \gls{eo} training datasets to ensure reliable performance of AI-based models in operational environments, as demonstrated in maritime surveillance applications.
\end{abstract}


\glsresetall

\section{Introduction} \label{sec:intro}

\gls{eo} satellites play a critical role in sustainability applications such as environmental monitoring and disaster response, as well as maritime surveillance, border security, and vehicle tracking \cite{leyva2023satellite}. The effectiveness of these applications depends on the ability to reliably extract information from satellite imagery, which in turn requires robust detection and classification algorithms. For instance, timely ship detection can assist in combating illegal fishing, piracy, and maritime accidents \cite{GALDELLI202515}; accurate monitoring of urban growth supports sustainable development policies \cite{HEIDARY20234272}; and rapid damage assessment after natural disasters enables efficient emergency response \cite{Zhang03052023}.

Despite their importance, \gls{eo} images are often degraded by physical phenomena such as cloud coverage, pointing errors, and atmospheric turbulence. While cloud coverage is frequently addressed in dataset preparation, turbulence-induced distortions and spacecraft-induced pointing jitter are largely neglected in existing datasets. This omission is problematic. Turbulence leads to spatial blurring and image distortions, while jitter causes misalignment and pixel misregistration—both of which can drastically reduce the performance of AI-based detection systems in real operational conditions. 



Most public \gls{eo} datasets used for training AI models assume clear and stable imaging conditions. Consequently, their performance may exhibit significant degradation when applied to real-world imagery affected by turbulence and jitter. This limits their applicability in mission-critical scenarios such as maritime surveillance, where false negatives in vessel detection may have severe economic, environmental, or security implications.




In this work, we build upon and extend the analysis presented in \cite{soret2024semantic}, where semantic communication concepts were introduced for \gls{eo} scenarios. Specifically, we further investigate the impact of atmospheric turbulence on optical satellite imagery and incorporate platform-induced pointing jitter to obtain a more realistic assessment of detection performance. By extending the Zernike-based turbulence simulator proposed in \cite{chimitt2020simulating}, we generate distorted images that better capture operational degradations. These synthetic yet realistic images can be used both for training and for evaluating AI-based detection models under realistic conditions.


As a case study, we apply our framework to the training and testing of a YOLOv8 vessel detection model \cite{yolov8_ultralytics}, and compare its performance against RetinaNet \cite{retinanet_pytorch}. YOLOv8 has been selected as it is a well-established detection algorithm that provides real-time inference with high accuracy and lower computational cost compared to other approaches such as RetinaNet and Mask R-CNN. Results show a substantial reduction in recall under turbulence and jitter scenarios, demonstrating the need to incorporate such physical effects in dataset generation. Beyond ship detection, the proposed approach can benefit a broad range of \gls{eo} applications, including disaster monitoring, border surveillance, and precision agriculture, by providing more robust AI-based analytics in degraded imaging environments.


The rest of this paper is organized as follows:  Section \ref{sec:sys} describes the considered \gls{eo} scenario. Section \ref{sec:img} details the adopted image-distortion model. Section \ref{sec:res} presents the performance evaluation and discusses the obtained results. Finally, Section \ref{sec:con} concludes the paper with final remarks.

\section{\gls{eo} scenario} \label{sec:sys} 
\begin{figure}
    \centering
    \includegraphics[width=\linewidth]{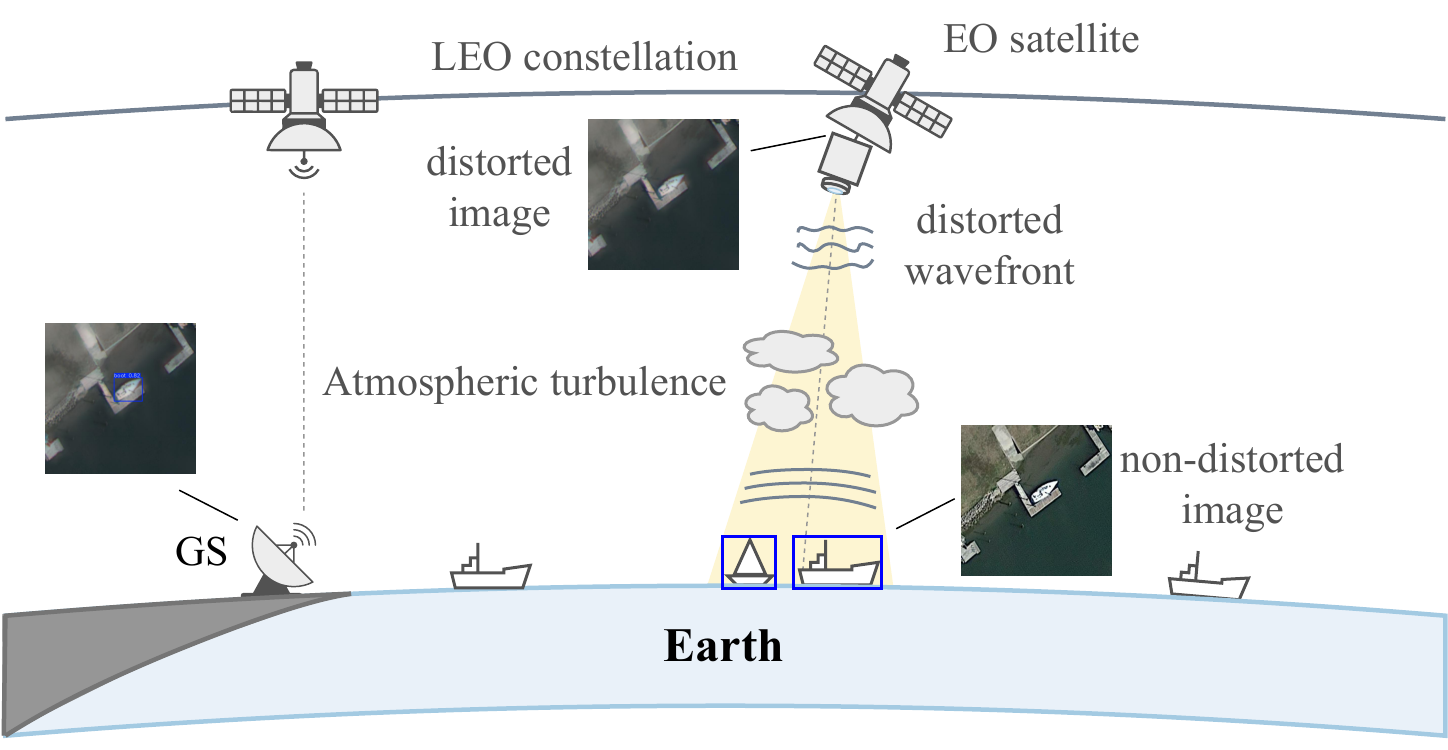}
    \caption{Scenario with a single LEO satellite and the presence of atmospheric turbulence.}
    \label{fig:DistorsionAtm}
\end{figure}

To illustrate our scenario in a realistic context, we consider a representative \gls{eo} mission where a \gls{leo} satellite captures high-resolution optical images of maritime regions. Such scenarios are of growing interest for vessel detection and tracking, illegal fishing prevention, and maritime safety operations. Figure \ref{fig:DistorsionAtm} shows the observation setup of a \gls{leo} satellite equipped with a camera. As the satellite moves along its orbit, it acquires optical images of the Earth's surface. Light reflected by objects on the Earth's surface propagates upward through the atmosphere, where refractive index fluctuations distort the wavefront and degrade spatial resolution. In parallel, the imaging system is subject to mechanical vibrations and attitude instabilities that introduce pointing jitter, further blurring the acquired images. The resulting measurements therefore deviate from the ideal scene, complicating object detection and classification.  

Once the images are captured, the satellite establishes a \gls{los} \gls{rf} link with a \gls{gs} to downlink the data. Object detection processing is assumed to be performed in the ground segment, with sufficient computational resources available to execute the required algorithms. 
This configuration enables a direct assessment of how physical disturbances and transmission effects influence the overall performance of the detection system.

The maximum achievable data rate for the \gls{rf} link at time $t$ depends on the received \gls{snr}, following Shannon capacity:
\begin{equation}
    R_{t} = B \log_{2}\left(1 + \hbox{SNR}_{t}\right),
    \label{eq:bitrate}
\end{equation}
where $B$ denotes the channel bandwidth. The instantaneous received \gls{snr} is given by $\hbox{SNR}_t =PG_{T}G_{R}/(L_{p}L_{b}\sigma^{2})$
where $P$ is transmit power; $G_{T}$ and $G_{R}$ are the transmitter and receiver parabolic antenna gains, respectively: $\sigma^{2}$ is the noise power; $L_{p}$ represent the pointing loss; and $L_{b} = \left(4\pi d f/c \right)^{2}$ is the free-space path loss, 
Here, $d$ is the instantaneous distance between the satellite and the \gls{gs}, $f$ is the carrier frequency, and $c$ is the speed of light. The distance $d$ is a function of the satellite orbital altitude and the elevation angle $\varepsilon$, and can be derived geometrically using the Pythagorean theorem.


The volume of data to be transmitted scales with the captured area, the \gls{gsd}, and the number of bits per pixel. For high-resolution optical payloads, this volume can easily exceed several hundred Mbps, making compression and on-board processing essential. In this context, capture conditions must be carefully considered, as they directly influence both algorithmic performance and transmission efficiency. More importantly, by modeling turbulence and jitter during the image acquisition stage, the proposed framework enables an end-to-end performance evaluation that accounts for both physical distortions and communication constraints in realistic \gls{eo} missions. Thus, we can establish a threshold of turbulence that guarantees a satisfactory outcome when processing the data at ground. 


\section{Image Distortion Model} \label{sec:img}
In satellite imaging systems, the fidelity of the captured scene is influenced by both the propagation medium and the stability of the platform. In particular, atmospheric turbulence and pointing jitter are two critical impairments that directly impact the \gls{gsd} and the spatial resolution of \gls{eo} images. Their effects, often neglected in existing datasets, are crucial to model when assessing the robustness of AI-based detection algorithms.
\subsection{Atmospheric Turbulence}


Atmospheric turbulence, which gives rise to the phenomenon known as atmospheric scintillation, is primarily caused by temperature--and pressure--induced fluctuations in the refractive index of air \cite{andrews2005laser}. These random variations distort both the amplitude and the phase of the received optical signal, making turbulence a fundamental limitation of FSO communication links \cite{sanchez2026turbulence}. In image systems, the same fluctuations distort the optical wavefront as light propagates through the atmosphere, resulting in scintillation, blur, and loss of contrast. The severity of these distortions depends on altitude, weather conditions, and time of day, and can be characterized by the scintillation index $\sigma_I^2$. For vertical ground-to-satellite propagation, the optical field is typically modeled as a spherical wave. In this case, the scintillation index is given by 
\begin{equation}
    \sigma_{I}^{2} = \exp{\left[\frac{0.49\sigma_{R}^{2}}{\left(1 + 0.56\sigma_{R}^{12/5}\right)^{7/6}} + \frac{0.51\sigma_{R}^{2}}{\left(1 + 0.69\sigma_{R}^{12/5}\right)^{5/6}}\right] - 1},
    \label{eq:varIrr}
\end{equation}
where $\sigma_{R}^{2}$ denotes the Rytov variance \cite{andrews2005laser}. Since atmospheric turbulence is predominantly located in the lower layers of the atmosphere,  $\sigma^2_R$ tends to saturate with altitude. The corresponding end-to-end variance is calculated as \cite{andrews2005laser}:
\begin{equation}
 \begin{split}
     \sigma_{R}^{2} = &2.25k^{7/6}(H - h_{0})^{5/6}\sec(\zeta_U)^{11/6} \times \\
     &\times \int_{h_{0}}^{H}C_{n}^{2}(h) \left(1 - \frac{h-h_{0}}{H-h_{0}}\right)^{5/6} \left(\frac{h-h_{0}}{H-h_{0}}\right)^{5/6}\, \mathrm{d}h,
  \end{split}
     \label{eq:varRytov}
\end{equation}
where $k=2\pi/\lambda$ represents the optical wave number, $H$ is the orbital altitude, $h_{0}$ is the height of the transceiver on the ground, $\zeta_{U}$ is the zenith angle of the propagation path and $C_{n}^{2}(h)$ denotes the refractive index structure parameter and describes turbulence strength at altitude $h$. Among existing models, the Hufnagel-Valley (HV) profile \cite{andrews2005laser} is widely used, 
as it incorporates altitude, wind speed, and ground-level conditions. In this work, the HV model has been adapted to include a variable $C_n^2(0)$, enabling the simulation of diurnal variations in turbulence intensity. Thus:
\begin{equation}
    \label{eq:HVmodel}
    \begin{split}
        C_{n}^{2}(h) = & 0.00594(w/27)^{2}(10^{-5}h)^{10}\exp{(-h/1000)} + \\
       & + 2.7 \times 10^{-16}\exp{(-h/1500)} + A\exp{(-h/100)},
    \end{split}
\end{equation}
with $w$ being the rms wind speed (pseudowind) in meters per second [m/s], and $A$ a nominal value of $C_{n}^{2}(0)$ at the ground in [m$^{-2/3}$].

To emulate image distortions, we build upon the Zernike-based turbulence simulator of Chimitt and Chan\cite{chimitt2020simulating}, which avoids expensive wave-propagation calculations by sampling intermodal and spatially correlated Zernike coefficients, the latter reflecting the physical properties of atmospheric turbulence. Each coefficient corresponds to a specific aberration (e.g., tilt, defocus, astigmatism) and is generated from a multivariate Gaussian distribution with covariance matrix $\mathbf{C}$, where the $(j,j')$-th element is given by:
\begin{equation}
    \label{eq:matrixC}
    [\mathbf{C}]_{j,j'} = \mathbb{E}[a_{j}^{*}a_{j'}].
\end{equation}
Here $\mathbb{E}[\cdot]$ denotes the expectation operator, $a_{j}$ and $a_{j'}$ are the Zernike coefficients, and $j$ indexes the Zernike mode.
The covariance matrix $\mathbf{C}$ captures the intermodal correlations between coefficients, ensuring that the sampled aberrations follow the statistical behavior of real atmospheric turbulence.


The distorted wavefront is then reconstructed as
\begin{equation}
    \label{zerkpol}
    \Phi(R\boldsymbol{\rho}) = \sum_{j=1}^{N} a_{j}Z_{j}(\boldsymbol{\rho})
\end{equation}
where $Z_{j}(\boldsymbol{\rho})$ denotes the $j$-th Zernike polynomial, $\boldsymbol{\rho} = [\rho, \theta]^{T}$ is the normalized polar coordinate vector with $0\leq\rho\leq1$ and $R\boldsymbol{\rho} = [R\rho, \theta]^{T}$ with $R = D/2$ being the aperture radius and $D$ the lens diameter. The Zernike coefficients depend on $D$, the refractive index structure parameter $C_{n}^{2}$ profile, the wavelength $\lambda$, and the focal length $f$. The distorted wavefront $\Phi(R\boldsymbol{\rho})$ is then converted into a \gls{psf}, which determines the effective blur applied to the image.




The simulator divides the input image into smaller blocks and generates the corresponding \gls{psf} for each block. A spatially varying blur is then applied to the blocks, and the tilt values are sampled according to the covariance matrix $\mathbf{C}$, warping the image based on turbulence-induced motion.

Using the Hufnagel-Valley model, different values of $C_{n}^{2}(h)$ are obtained for the propagation path.. Following \cite{andrews2005laser}, and after some mathematical manipulation, these values can be averaged as:
\begin{equation}
    \label{eq:Cn2average}
    \begin{split}
        <C_{n}^{2}> = & \left(\frac{4.5}{H-h_{0}}\right) \Delta h \times \\ 
        & \times \sum_{h=h_{0}}^{H}\left(C_{n}^{2}(h) \left(\frac{h-h_{0}}{H-h_{0}}\right)^{5/6} \left(\frac{H-h}{H-h_{0}}\right)^{5/6}\right), 
    \end{split}
\end{equation}
where $\Delta h$ denotes the altitude step used to compute the average. The summation is evaluated over the discrete set of altitudes $h = [h_{0}:\Delta h:H]$ and the resulting average $<C_{n}^{2}>$ for the vertical path is then used as input to the simulator.

\begin{figure}
    \centering
    \includegraphics[width=\linewidth]{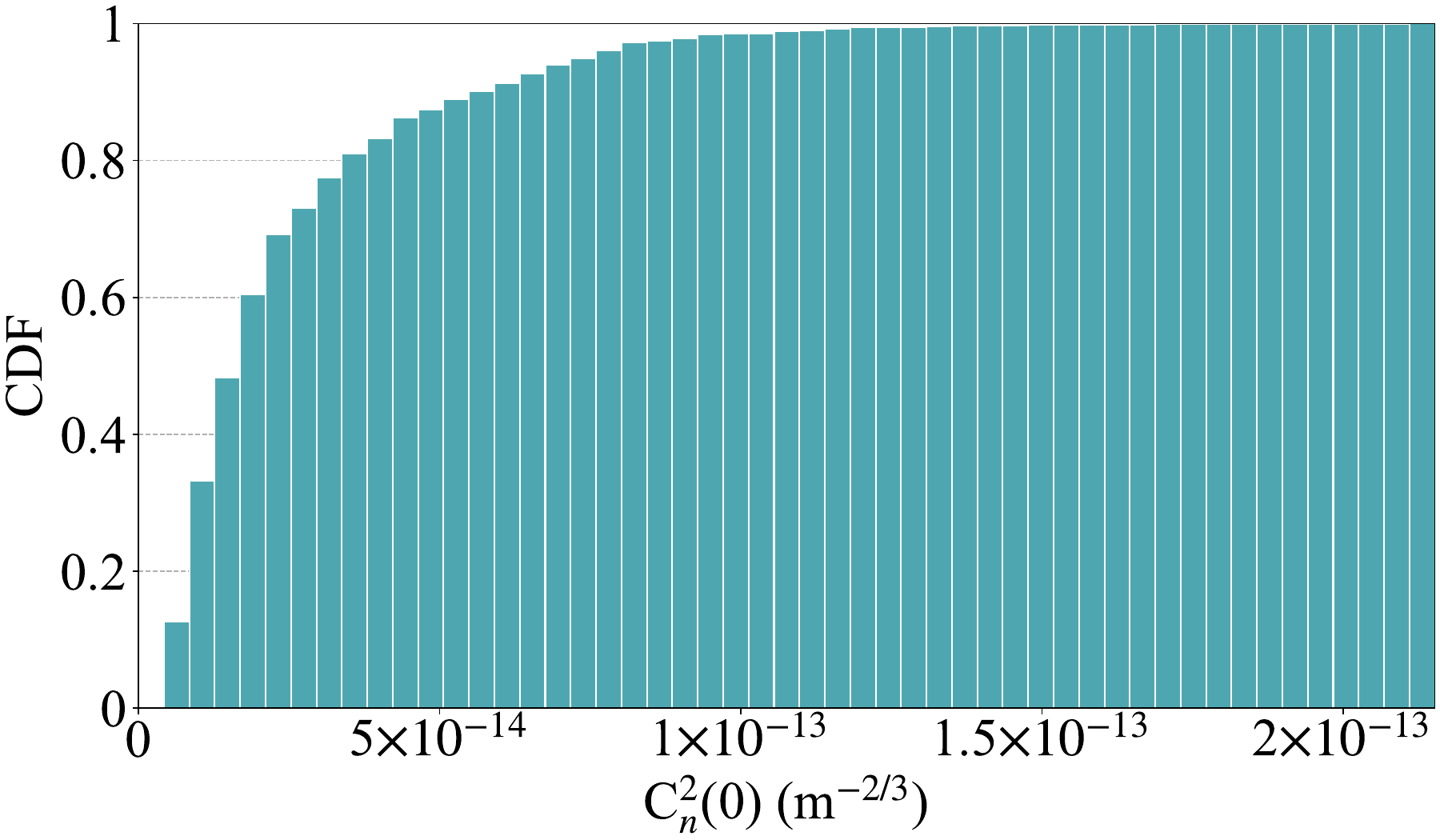}
    \caption{\gls{cdf} of the refractive index structure parameter at sea level $C_{n}^{2}(0)$.} 
    \label{fig:cdf}
    \vspace{-0.4cm}
\end{figure}

This framework allows the generation of physically grounded distortions consistent with atmospheric conditions at different altitudes and times of day. Compared to prior work, our contribution lies in extending the simulator to vertical paths, 
incorporating an altitude-dependent refractive-index structure parameter $C_n^2(h)$, and tailoring the model to EO imagery, where robustness of AI-based object detection is the primary concern. As an illustrative example, experimental ground-level measurements are used to define a \gls{cdf} for $C_{n}^{2}(0)$, as shown in Figure \ref{fig:cdf} highlighting the variability of this parameter. In our framework, this variability is captured by allowing $C_n^2(0)$$-$and thus the corresponding $C_n^2(h)$ profile$-$to vary with the satellite’s orbital position, yielding position-dependent blur and a more realistic representation of atmospheric effects.

\subsection{Pointing jitter in optical satellite systems}

In addition to atmospheric effects, high-resolution EO systems are highly sensitive to pointing jitter. In satellite imaging systems, jitter refers to rapid, random time-varying deviations in the orientation of the sensor's line of sight, typically caused by mechanical vibrations of the platform,  attitude control system inaccuracies, residual tracking noise or photoelectric sensor fluctuations. These perturbations lead to angular errors that misalign the beam with respect to its intended direction \cite{Zaman2020}, being particularly relevant in \gls{leo} satellites (inter-satellite or satellite-to-ground laser links), where high-resolution imaging demands precise attitude stability to maintain image quality. Even small angular deviations can lead to pixel misregistration and effective blurring, particularly when the displacement exceeds the \gls{gsd}.

Although satellite jitter may contain temporal correlation, in this work it is modeled per image as a zero-mean Gaussian angular displacement with respect to the camera boresight. 
This choice is justified because the analysis relies on the marginal distribution of pointing errors 
and the dataset is treated as a collection of independent images (i.e., no temporal sequences are exploited), so inter-frame correlation is neglected. 
Following \cite{Liu2021}, angular deviations $\theta_x$ and $\theta_y$ are modeled as independent Gaussian random variables. Their respective probability density functions (PDFs) are given by:
\begin{subequations}
    \begin{equation}
    f(\theta_x) = \frac{1}{\sqrt{2\pi} \sigma_x} \exp\left(-\frac{\theta_x^2}{2\sigma_x^2}\right), 
    \end{equation}
    \begin{equation}
    f(\theta_y) = \frac{1}{\sqrt{2\pi} \sigma_y} \exp\left(-\frac{\theta_y^2}{2\sigma_y^2}\right)
    \end{equation}    
    \label{eq:angularPE}
\end{subequations}
where $\sigma_x$ and $\sigma_y$ are the standard deviations of the angular deviations along each axis. When the standard deviations of the azimuth and elevation angles are equal, the radial displacement follows a Rayleigh distribution with deviation $\sigma_r$ \cite{BadiaAldecoa2024}. In our framework, these deviations are incorporated directly into the Zernike covariance matrix $\mathbf{C}$, thus unifying turbulence and jitter within a single statistical model.

\section{Evaluation and results} \label{sec:res}
To evaluate the impact of atmospheric turbulence and pointing jitter on \gls{eo}-based object detection, we trained and tested the YOLOv8 and RetinaNet algorithms \cite{yolov8_ultralytics, retinanet_pytorch} for vessel detection. The dataset used for this purpose is the \textit{Ships-Google-Earth} dataset \cite{Ships_Google_Earth_Dataset}, consisting of 794 images from \textit{Google Earth}, all resized to 600$\times$600 pixels. Data augmentation and dataset split follow the configuration described in \cite{Ships_Google_Earth_Dataset}. We used the YOLOv8m model, pretrained on the COCO dataset, and the retinanet\_resnet50\_fpn\_v2 model for RetinaNet, also pretrained on COCO. Both models were trained for 100 epochs with a batch size of 16. Images were distorted using the extended Zernike-based simulator described in Section~\ref{sec:img}, with parameter settings consistent with the WorldView-3 satellite \cite{worldview-13} and RF link parameters taken from \cite{ngeo2022}. A summary of the system parameters used in this work is provided in Table \ref{tab:parameters}. 
Each RGB channel was independently processed through the simulator and recombined to generate the final distorted image. 

The detection performance was quantified using the recall metric, which measures the ability of a model to identify all relevant cases 
in the dataset, defined as: 
\begin{equation}
    \hbox{Recall} = \frac{\hbox{Number of detected vessels}}{\hbox{Actual number of vessels}},
\end{equation}
showing the sensitivity of the model to correctly detect vessels present in the images. 

\renewcommand{\arraystretch}{1.2}
\begin{table}
    \vspace*{2mm}
    \centering
    \begin{tabular}{lcc}
        \hline
        \textbf{Parameter} & \textbf{Symbol} & \textbf{Value} \\
        \hline
        RGB wavelengths (nm) & $\lambda_{R,G,B}$ & 650, 520, 440 \\
        \hline
        Ground stations height (m) & $h_{0}$ & 0 \\
        LEO orbit height (km) & $H$ & 617 \\
        Ground sample distance (m/pixel) & $GSD$ & 0.5 \\
        Lens diameter (m) & $D$ & 1.1 \\
        Focal length (m) & $f$ & 16 \\
        Wind speed (m/s) & $w$ & 21 \\
        Pointing jitter ($\mu$rad) & $\sigma_{r}$ & 5, 10, 20, 30, 40, 50 \\
        \hline
        Carrier frequency (GHz) & $f$ & 20 \\
        Bandwidth (MHz) & $B$ & 500 \\
        Transmission Power (W) & $P$ & 10 \\
        Noise power (dBW) & $\sigma^{2}$ & $-$117.77 \\
        Antenna Gain (dBi) & $G_T-G_R$ & 32.13 $-$ 34.20 \\
        Pointing Loss (dB) & $L_{p}$ & 0.3 \\
        Antenna efficiency ($-$) & $\eta$ & 0.55 \\
        Elevation angle ($^{\circ}$) & $\varepsilon$ & 25 \\
        \hline
        
        \hline
    \end{tabular}
    \caption{System parameters.}
    \vspace*{-0.3cm}
  \label{tab:parameters}
\end{table}

\subsection{Impact of Atmospheric Turbulence}
The strength of turbulence can be parameterized through the scintillation index, defined as the normalized irradiance variance, $\sigma_{I}^{2}$. Weak-turbulence regimes correspond to values $\sigma_{I}^{2} \ll 1$, whereas moderate and strong turbulence are typically associated with $\sigma_{I}^{2} \sim 1$ and $\sigma_{I}^{2} > 1$, respectively \cite{andrews2005laser}. It should be noted that, for horizontal propagation paths, values of $C_{n}^{2}(0)$ above 10$^{-13}$ m$^{-2/3}$ generally lead to irradiance variances $\sigma_{I}^{2} > 1$. In contrast, for vertical propagation, the effective $C_{n}^{2}$ decreases significantly with altitude, as atmospheric conditions gradually approach free-space conditions. 
The corresponding irradiance-variance values for the turbulence scenarios considered in the simulations are summarized in Table~\ref{tab:Turbulencia}.


\renewcommand{\arraystretch}{1.2}
\begin{table}[t]
    \vspace*{2mm}
    \centering
    \begin{tabular}{lcc}
        \hline
        {\bf $\boldsymbol{C_{n}^{2}}$(0) (m$\boldsymbol{^{-2/3}}$)} & {\bf Average $\boldsymbol{C_ {n}^{2}}$ (m$\boldsymbol{^{-2/3}}$)} & {\bf $\boldsymbol{\sigma_{I}^{2}}$}\\
        \hline
        4.7 $\times$ 10$^{-17}$ & 4.9269 $\times$ 10$^{-20}$ & 0.0524\\
        5 $\times$ 10$^{-15}$ & 5.0140 $\times$ 10$^{-20}$ & 0.0533 \\
        1.7 $\times$ 10$^{-14}$ & 5.2009 $\times$ 10$^{-20}$ & 0.0556\\
        8.3 $\times$ 10$^{-13}$ & 1.7821 $\times$ 10$^{-19}$ & 0.2071\\
        2 $\times$ 10$^{-12}$ & 3.6080 $\times$ 10$^{-19}$ & 0.4138\\
        5 $\times$ 10$^{-12}$ & 8.2796 $\times$ 10$^{-19}$ & 0.8460\\
        \hline        
    \end{tabular}
    \caption{Values of the irradiance variance $\sigma_{I}^{2}$ as a function of the structure parameter $C_{n}^{2}$ at sea level \cite{maharjan2022atmospheric}.} 
    \label{tab:Turbulencia}
\end{table}

\begin{figure}[t]
    \centering
    \begin{subfigure}[b]{0.24\textwidth}
        \centering
        \includegraphics[width=0.95\textwidth]{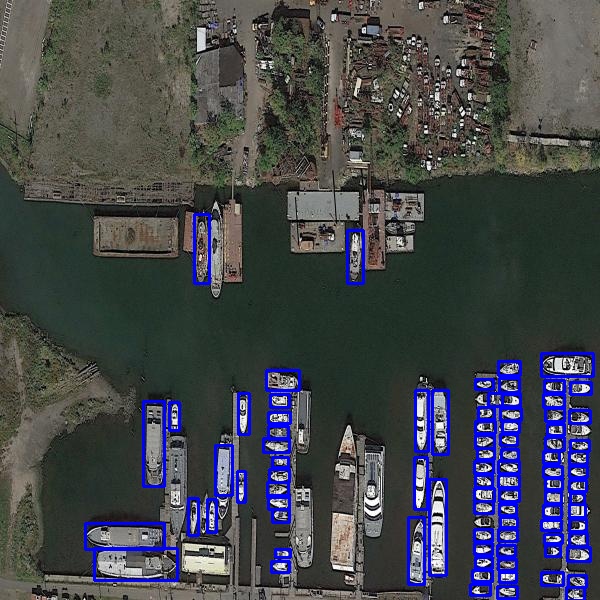}
        \caption{}
        \label{fig:nojitternoturb}
    \end{subfigure}
    \hfill
    \begin{subfigure}[b]{0.24\textwidth}
        \centering
        \includegraphics[width=0.95\textwidth]{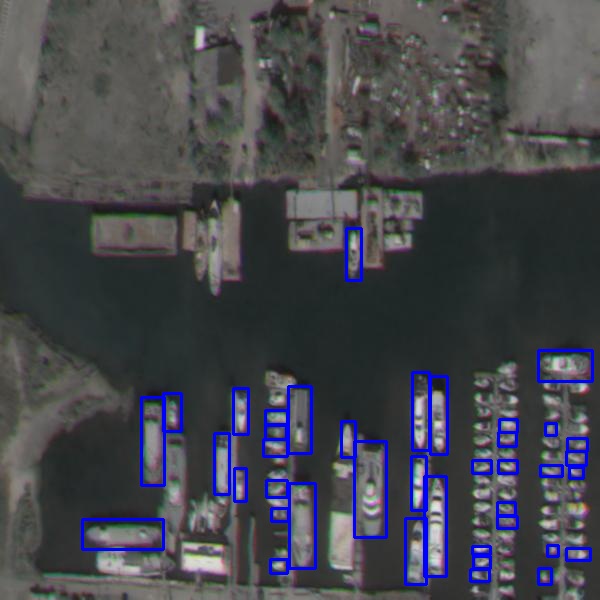}
        \caption{}
        \label{fig:nojitter}
    \end{subfigure}
    \vspace{0.5em}
    \begin{subfigure}[b]{0.24\textwidth}
        \centering
        \includegraphics[width=0.95\textwidth]{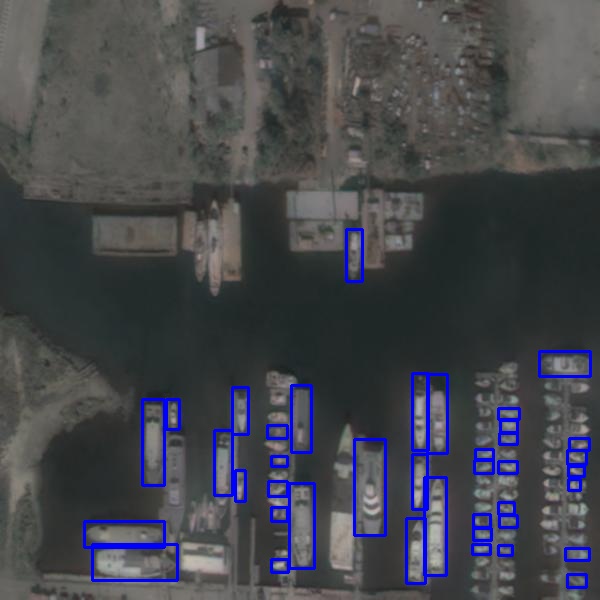}
        \caption{}
        \label{fig:jitter1}
    \end{subfigure}
     \hfill
    \begin{subfigure}[b]{0.24\textwidth}
        \centering
        \includegraphics[width=0.95\textwidth]{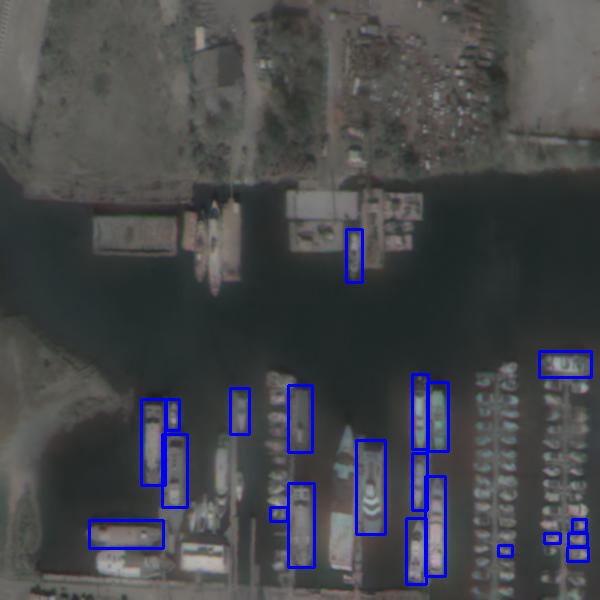}
        \caption{}
        \label{fig:jitter2}
    \end{subfigure}
    \hfill
    \caption{Impact of the atmospheric turbulence in ship detection and localization using YOLOv8 with (a) no turbulence and no jitter, (b) $\sigma_{I}^{2} =$ 0.0524 and no jitter, (c) $\sigma_{r} =$ 20 $\mu$rad and no turbulence, and (d) $\sigma_{I}^{2} =$ 0.2071 and $\sigma_{r} =$ 20 $\mu$rad.} 
    \label{fig:turbulentscenarios}
\end{figure}


Fig. \ref{fig:turbulentscenarios}(a) and \ref{fig:turbulentscenarios}(b) illustrate how atmospheric turbulence affects both the final image quality and the ship detection and localization results for a representative scenario. In this case, a very weak turbulence condition is considered, corresponding to a ground-level value $C_n^2(0)=4.7\times 10^{-17}$ m$^{-2/3}$ and YOLOv8 is employed as the detection model. Notably, even under this very weak turbulence level, Fig. \ref{fig:turbulentscenarios}(b) (with turbulence) shows visible image distortion and a clear degradation in vessel-detection performance when compared with Fig. \ref{fig:turbulentscenarios}(a) (without turbulence).

\begin{figure}[!t]
    \centering
    \includegraphics[width=1.07\linewidth]{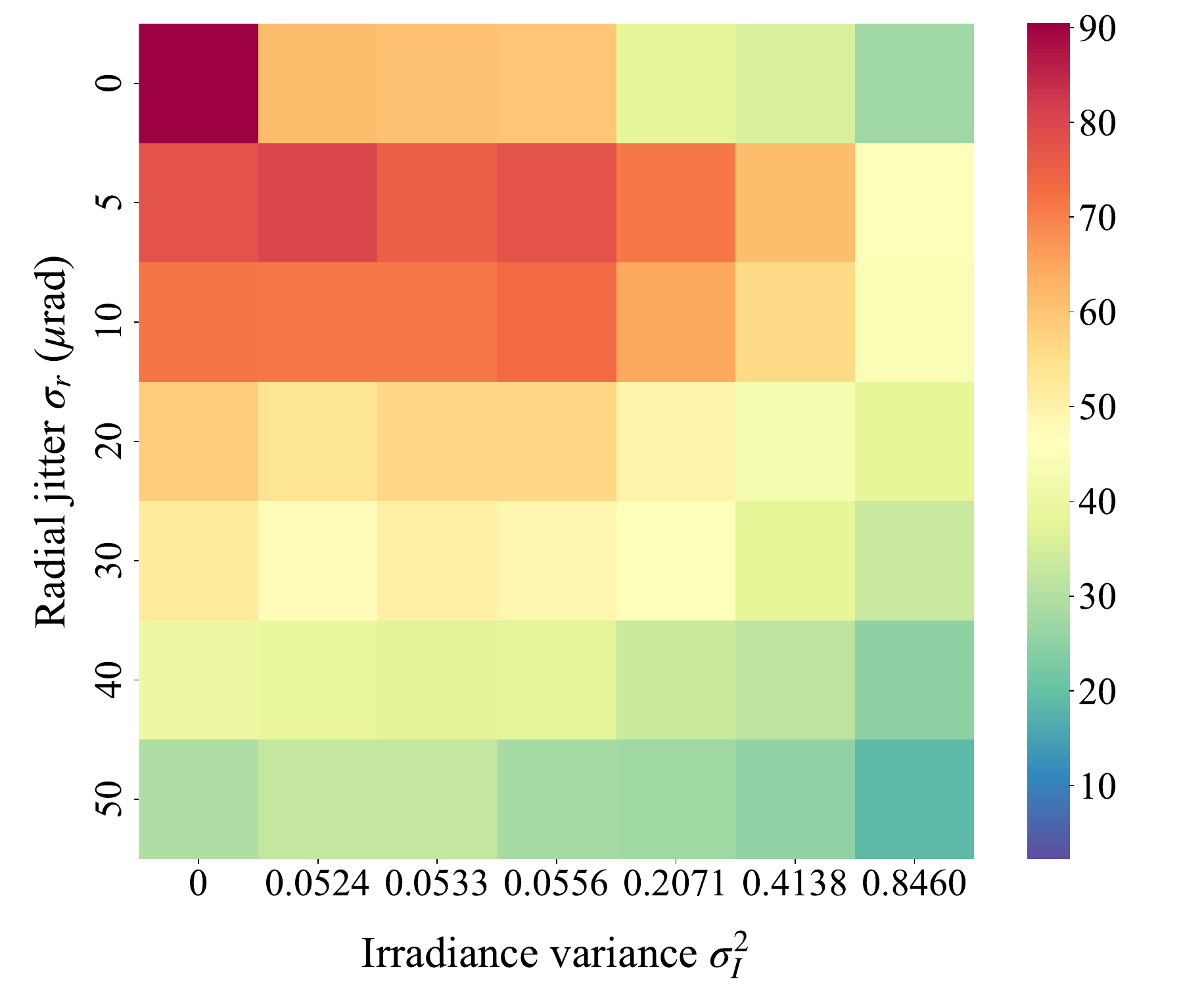}
    \caption{YOLOv8 recall over the irradiance variance $\sigma_{I}^{2}$ and the pointing radial jitter $\sigma_{r}$.}
    \vspace{-.2cm}
    \label{fig:recall-3d}
\end{figure}

\begin{figure}[!t]
    \centering
    \includegraphics[width=1.07\linewidth]{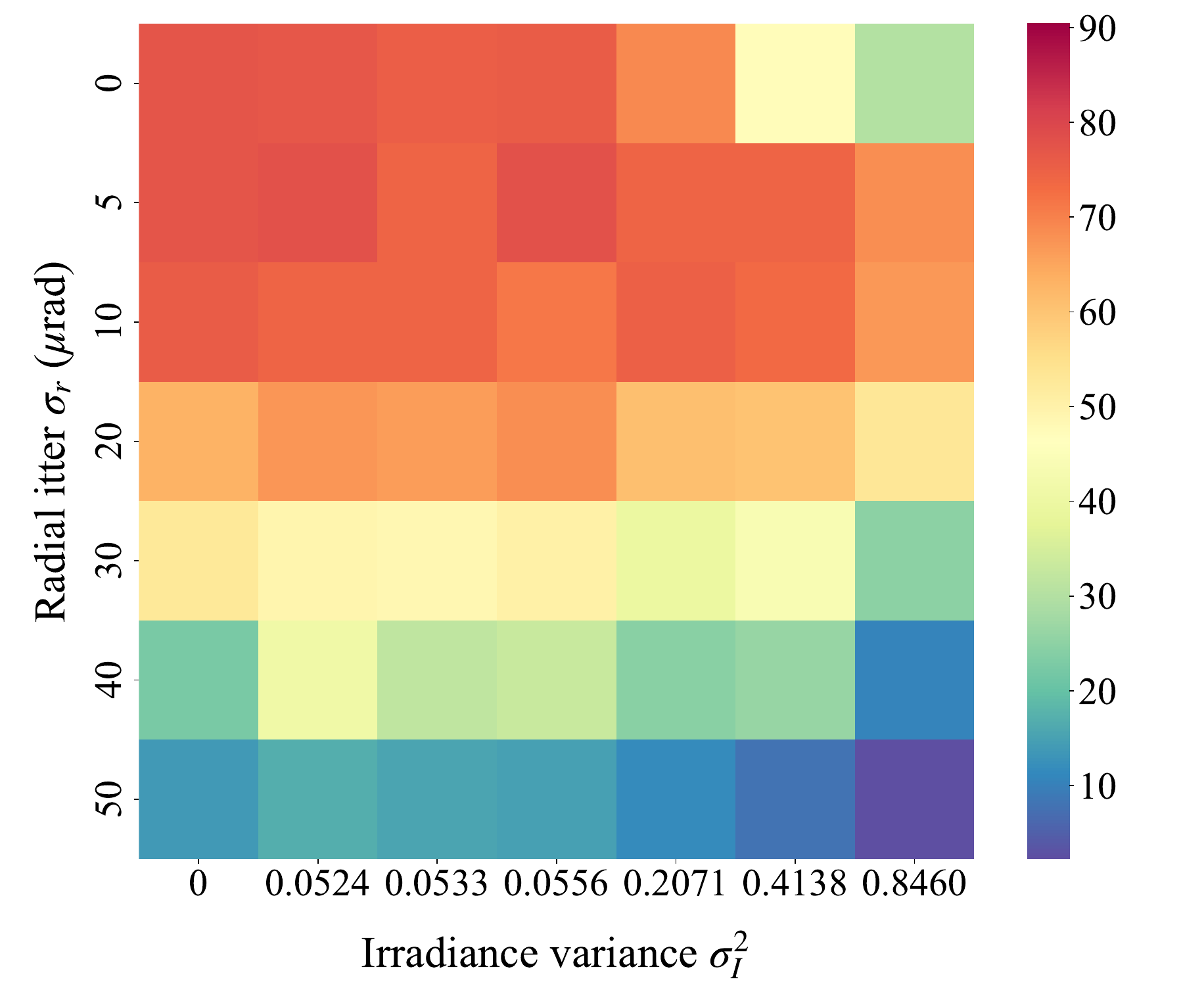}
    \caption{RetinaNet recall over the irradiance variance $\sigma_{I}^{2}$ and the pointing radial jitter $\sigma_{r}$.}
    \vspace{-.2cm}
    \label{fig:recall-RN}
\end{figure}

Fig.~\ref{fig:recall-3d} and Fig.~\ref{fig:recall-RN} quantify this effect in terms of recall for YOLOv8 and RetinaNet, respectively, across different turbulence regimes. As shown in Fig.~\ref{fig:recall-3d}, and consistent with the qualitative behavior observed in Fig.~\ref{fig:turbulentscenarios}, even a very weak turbulence regime with scintillation index $\sigma_{I}^{2}=0.0524$, corresponding to $C_n^2(0)=4.7\times 10^{-17}$ m$^{-2/3}$, causes a marked reduction in YOLOv8 recall, which drops from 90.42$\%$ in the absence of turbulence to 61.17$\%$. Fig.~\ref{fig:recall-RN} indicates that RetinaNet exhibits a more robust response under weak turbulence: although its recall under ideal conditions (77.08$\%$) is lower than that of YOLOv8, under weak turbulence it decreases only slightly to 76.79$\%$. For more adverse turbulence regimes, both detection models experience further performance degradation. Specifically, for a weak-to-moderate regime with scintillation index $\sigma_{I}^{2}=0.2071$, and a moderate-to-strong regime with $\sigma_{I}^{2}=0.846$, the recall of YOLOv8 decreases to 59.57$\%$ and 38.30$\%$, respectively, whereas RetinaNet attains 75.89$\%$ and 68.75$\%$, respectively. Overall, these results confirm that atmospheric turbulence, even in very weak regimes, can have a pronounced impact on detection performance.

\subsection{Impact of Pointing Jitter}

Fig. \ref{fig:turbulentscenarios}(a) and Fig. \ref{fig:turbulentscenarios}(c) illustrate the impact of radial pointing jitter $\sigma_r$ on the resulting image quality and on ship detection and localization for the same scenario analyzed in the previous subsection. In this example, a jitter standard deviation of $\sigma_r=20$ $\mu$rad is assumed and YOLOv8 is used as the detection model. As observed in Fig. \ref{fig:turbulentscenarios}(c), pointing jitter visibly blurs the image and introduces ship-detection errors that do not appear in the ideal, jitter-free case shown in Fig. \ref{fig:turbulentscenarios}(a). Moreover, this degradation becomes more pronounced when jitter is combined with atmospheric turbulence, as illustrated in Fig. \ref{fig:turbulentscenarios}(d).

Figs.~\ref{fig:recall-3d} and \ref{fig:recall-RN} further quantify the effect of radial pointing jitter in terms of recall for YOLOv8 and RetinaNet, respectively. In these experiments, the jitter standard deviation is varied from 0 to 
50 $\mu$rad, and the combined impact of atmospheric turbulence and jitter on both detection models can also be observed. When a small amount of jitter is present, part of the light scattering caused by turbulence is counteracted by that induced by pointing jitter, partially restoring the original alignment. For both YOLOv8 and RetinaNet, performance degradation becomes critical when jitter displacements exceed the \gls{gsd}, as object boundaries blur beyond recognition.


\subsection{Impact on Communication} 
Assuming the maximum imaging capacity specified in the WorldView-3 datasheet \cite{worldview-13}, and considering a GSD of 0.5~m/pixel and an 8-bit color depth, the total data volume captured in one day is approximately 7.42~TB, while the data acquired per orbital period amounts to 512~GB, requiring 15 orbital passes to complete. Conversely, using \eqref{eq:bitrate} and assuming that the \gls{gs} is located along the satellite’s ground track, the average downlink bitrate is 1.96~Gbps within a 4-minute visibility window, requiring approximately eight orbital passes to transmit all the acquired data.

Based on the YOLOv8 performance, since the recall drops below 60\% for $C_n^2(0) > 2\times10^{-14}$~m$^{-2/3}$, the transmission volume can be reduced by approximately 40\% by discarding images acquired under poor atmospheric conditions, thereby lowering the transmission requirement to five orbital passes. In contrast, RetinaNet maintains a recall above 75\% under the same conditions, which does not translate into a significant reduction in transmitted data. For YOLOv8, maintaining a threshold of $C_n^2(0) > 5\times10^{-15}$~m$^{-2/3}$ would reduce the transmitted data by up to 94\%, whereas setting it at $C_n^2(0) > 4\times10^{-14}$~m$^{-2/3}$ would only yield an 18\% reduction. Furthermore, this threshold can serve as a decision rule for image acquisition, optimizing on-board resources and storage efficiency.

\subsection{Discussion}

The results demonstrate a clear vulnerability of state-of-the-art AI detection models to realistic physical degradations. While YOLOv8 maintains recall above 90\% under ideal conditions, ensuring reliable positive detections, its recall drops drastically in the presence of atmospheric turbulence and pointing jitter, reducing sensitivity to objects of interest. In contrast, RetinaNet exhibits more consistent performance across degraded conditions but achieves lower recall under ideal scenarios. For maritime surveillance applications, this behavior translates into potentially missing up to 25\% and 40\% of vessels using RetinaNet and YOLOv8, respectively, under moderate turbulence conditions, which would be unacceptable in mission-critical scenarios.

These findings highlight the necessity of including turbulence and jitter effects when generating training datasets for AI-based \gls{eo} applications. By doing so, models can be made more robust to realistic observation conditions, improving their applicability not only to ship detection but also to urban monitoring, border control, and disaster response.

\section{Conclusion} \label{sec:con}

This paper has evaluated the combined impact of atmospheric turbulence and satellite-induced pointing jitter on Earth observation imagery, with a particular focus on vessel detection. To this end, the Zernike-based turbulence simulator in  \cite{chimitt2020simulating} was extended to vertical ground-to-satellite paths and incorporated jitter effects, enabling the generation of distorted imagesthat more realistically reflect operational conditions. The evaluation was carried out under different turbulence and jitter regimes. Results obtained with YOLOv8 show that both impairments can significantly degrade recall, whereas RetinaNet exhibits lower performance under ideal conditions but greater robustness to moderate variations in turbulence and jitter. These findings expose an important vulnerability of AI-based detection systems: their ability to detect all relevant targets degrades sharply as physical impairments increase. This has direct implications for mission-critical \gls{eo} applications such as maritime surveillance, border security, and disaster response, where missed detections may lead to severe operational risks. The proposed framework provides a practical tool for generating synthetic yet realistic training and evaluation datasets, paving the way toward more robust AI models for \gls{eo} tasks. 


Future work will extend this approach by incorporating additional impairments (e.g., sensor noise, optical misfocus) and by retraining detection models with turbulence- and jitter-aware datasets to improve resilience. The proposed methodology is broadly applicable beyond ship detection, offering benefits for urban change monitoring, agricultural analysis, and rapid disaster assessment.

\balance

\bibliographystyle{IEEEtran}
\bibliography{refs}


\end{document}